\documentclass[%
 aip,
 jmp,%
 amsmath,amssymb,
 reprint,%
]{revtex4-1}

\usepackage{subfigure}
\usepackage{graphicx}
\graphicspath{{./}{./images/}{./images_all/}}
\DeclareGraphicsExtensions{.png}

\usepackage{dcolumn}
\usepackage{bm}

\begin{document}

\preprint{AIP/123-QED}

\title[APPLIED PHYSICS LETTERS xx (2011)]{An ultrafast image recovery and 
recognition system implemented with nanomagnets possessing biaxial
magnetocrystalline anisotropy}

\author{Noel D'Souza}
\email{dsouzanm@vcu.edu.}
\author{Jayasimha Atulasimha}
\affiliation{Department of Mechanical and Nuclear Engr., Virginia Commonwealth University, Richmond,~VA~23284,~USA.}
\author{Supriyo Bandyopadhyay}
\affiliation{Department of Electrical and Computer Engr., Virginia Commonwealth University, Richmond,~VA~23284,~USA.}

\date{\today}

\begin{abstract}
A circular magnetic disk with biaxial magnetocrystalline anisotropy has four 
stable magnetization states which can be used to encode a pixel's shade in a 
black/gray/white image. By solving the Landau-Lifshitz-Gilbert equation, 
we show that if moderate noise deflects the magnetization slightly from a stable state, 
it always returns to the original state, thereby automatically de-noising
the corrupted image. The same system can compare a noisy input image with a stored 
image and make a matching decision using magneto-tunneling junctions. 
These tasks are executed at ultrahigh speeds ($\sim$2 ns for a 
512$\times$512 pixel image).
\end{abstract}

\maketitle

Consider a single crystal single-domain nanomagnet in the shape of a circular disk that has no 
in-plane shape anisotropy but has biaxial magnetocrystalline anisotropy. This magnet has four 
stable magnetization directions (`up', `right', `down', `left') as shown in Fig. 1 (a) \cite{noel}.

\begin{figure}

\includegraphics[width=2.5in]{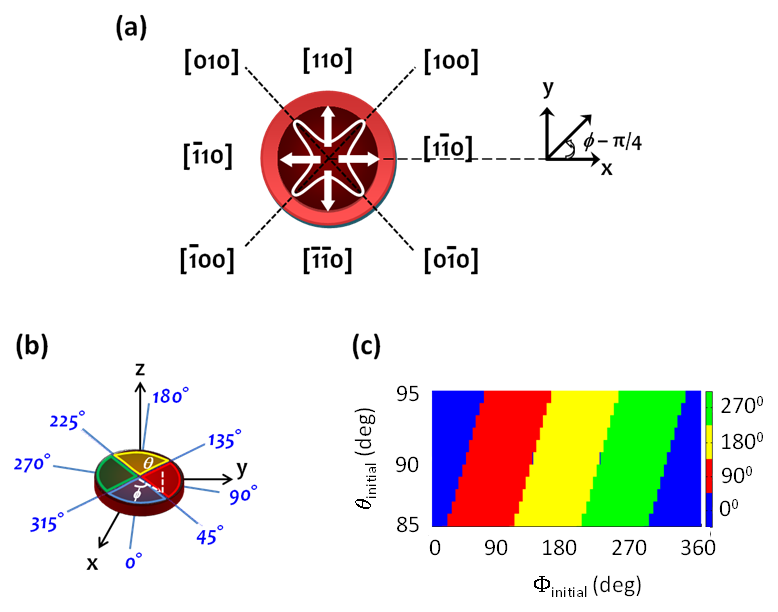}
\caption{(a) A circular disk-shaped single-crystal Ni layer with biaxial anisotropy, 
creating four possible magnetization directions or `easy' axes -- `up', `down', `left' and 
`right'. The energy landscape is depicted by the saddle-shaped curve, resulting in the four 
energy minima along the  axes. 
(b) The spherical coordinate system used to define the magnetization vector and the color scheme 
used to designate quadrants. (c) Color plot of the magnetization vector's final state for all 
allowed values of the initial state $\left (\theta_{initial},\phi_{initial} \right)$.}
\end{figure}

An array of such magnets can store a black-gray-white image by
storing the shade of every pixel
 in one of the four stable states of a magnet. If corrupted by moderate noise,
 the magnetization dynamics automatically recover the stored image,
 so that there is built-in error correction. When integrated with magneto-tunneling 
 junctions, the same array can recognize 
images by comparing them with stored images pixel by pixel. These tasks are executed
with ultrahigh speed since no software is needed.

The four states of the magnet  encode black, gray, white and gray. 
We could have encoded two different shades of gray in the four states, but the 3-shade scheme is simpler. 
Noise can corrupt the stored image by perturbing the magnetization vector and deflecting
it away from the initial stable state. The ensuing magnetization dynamics is studied using 
the Landau-Lifshitz-Gilbert (LLG) equation to determine the final state of every perturbed 
magnet (or pixel). For this purpose, we assume that the magnets are spaced far enough apart 
that dipole interaction between them can be ignored. Since the useful pixel density in any 
optical image is limited by the wavelength of light (0.5 - 1.0 $\mu$m for visible light), we 
can space the magnets apart by 1 $\mu$m, which makes the inter-magnet dipole interaction 
negligible.  Using knowledge of the magnetization dynamics of such a 4-state magnet, 
we demonstrate image recovery in a 512$\times$512 pixel black-white-gray image in $\sim$2 ns 
while conventional field programmable gate array (FPGA)- and CMOS-based image filters would 
have taken several  $\mu$s \cite{kawai}. 

In order to demonstrate image recovery, we consider an isolated circular nanomagnet 
 of nickel with diameter 100 nm and thickness 10 nm. Nickel has an `easy' axis of magnetization along 
the $\langle 111 \rangle$ direction, a `medium' axis along the $\langle 110 \rangle$ direction
 and a `hard' axis along the $\langle 100 \rangle$ direction. We assume the nickel layer to be 
 in the (001) plane. In this  plane, the `easy' axis is the $\langle 110 \rangle$ direction 
 (`medium' for the entire crystal) while the `hard' axis is along the $\langle 100 \rangle$ 
 direction. Because the thickness of the magnet is ten times smaller than the diameter, 
 out-of-plane excursion of the magnetization vector is energetically costly, 
 albeit not impossible. 

As illustrated in Fig. 1 (a), the easy axes of a single-crystal Ni nanomagnet in the x-y 
plane (magnet's plane) are along the [110], [$\overline{1}~\overline{1}$0], 
[$\overline{1}$10] and [1$\overline{1}$0] 
directions, in Miller notation. The saddle-shaped curve and arrows depict the potential 
energy landscape and stable directions of the magnetization vectors in the four (degenerate) 
minimum energy locations. The coordinate axis system is rotated by 45$^{\circ}$ 
in the magnet's plane (i.e. $\phi \rightarrow \phi - \pi/4$), where $\phi$ is the 
angle subtended by the in-plane component of the magnetization vector with the +x-axis.
This rotational transformation ensures that the energy minima are along the $\pm$x-axes 
($\phi$ = 0$^{\circ}$ or $\pm180^{\circ}$) and $\pm$y-axes ($\phi = \pm90^{\circ}$). 
The angle subtended by the magnetization vector with the z-axis is $\theta$.
 When $\theta = 90^{\circ}$, the magnetization vector is in the plane of the magnet.

If noise deflects the magnetization vector of a magnet away from its initial stable state 
to a new state, 
the latter evolves with time in accordance with the Landau-Lifshitz-Gilbert (LLG) equation. 
We take the new state as the initial condition and 
solve the LLG equation to determine the final state. We find that 
as long as the noise amplitude is not large enough to deflect the magnetization
vector closer to another stable orientation or cause large out-of-plane excursion, 
the vector always returns to the 
initial stable orientation. This happens relatively fast in a few ns (for realistic parameters).
Thus, the system acts like {\it associative memory} and could be used in that role as well.
However, here we study image recovery.

The total magnetic energy of the single-domain nanomagnet per unit voulme is
\begin{multline}
E_{total} (t)  =  E_{magnetocrystalline} (t) + E_{shape} (t) 	\\
 =  {{K_1}\over{4}} sin^2 2 \theta(t)+ \left[ {{K_1}\over{4}} sin^4 \theta(t)
+{{K_2}\over{16}} sin^2 \theta(t) sin^2 2 \theta(t) \right  ]  \\
 +{{\mu_0}\over{2}} M_s^2 \left [N_z cos^2 \theta(t)+ N_x sin^2 \theta(t) \right ] ,
\label{energy-eq}
\end{multline}
where $K_1$ and $K_2$ are the first and second order magnetocrystalline anisotropy constants, 
$\mu_0$ is the permeability of free space, $M_s$ is the 
saturation magnetization and $N_x, N_y, N_z$ are the components of the demagnetization factors 
along the x-, y-, and z-axes, respectively. Since we consider a circular nanomagnet,
$N_x = N_y$. These demagnetizing factors are determined from the oblate spheroid estimation
\cite{cullity}.

Since the magnet is single-domain, the magnetization $\mathbf{M}$ has a constant magnitude. 
This allows us to define a unit vector $\mathbf{n_m} = \mathbf{M}/|\mathbf{M}|= \mathbf{e_r}$
in the radial direction. The magnetization at any instant is therefore uniquely specified by 
 $\theta(t)$ and $\phi(t)$.

In any arbitrary orientation [$\theta(t), \phi(t)$], the magnetization vector experiences
a torque per unit 
volume due to magnetocrystalline and shape anisotropy. The latter can be written as 
\begin{equation}
\mathbf{T_E} (t)=-\mathbf{n_m} \times \mathbf{\nabla}E_{total} (t)=-\mathbf{e_r} \times
\mathbf{\nabla}E_{total} \left (\theta(t),\phi(t) \right ) ,
\end{equation}
where $E_{total}(t)$ is given by Equation (\ref{energy-eq}). It can be verified that 
the torque vanishes when $\theta = \pi/2$ and $\phi = n \pi/2$ [$n$ = 0, 1, 2, 3], indicating 
that these are the four stable orientations. Whenever the magnetization is deflected from a 
stable orientation by noise, it experiences the above restoring torque that attempts to return 
the magnetization to the initial state. 

In order to study the restoring dynamics, we solve the Landau-Lifshitz-Gilbert (LLG) equation
\begin{equation}
{{d \mathbf{n_m}}\over{dt}} + \alpha \left ( \mathbf{n_m} (t) \times {{d\mathbf{n_m} (t)}
\over {dt}} \right )= {{2 \mu_B}\over{\hbar M_s \Omega}} \mathbf{T_E}  (t) ,
\end{equation}
where $\alpha$ is the Gilbert damping coefficient (0.045 for Ni), 
$\Omega$ is magnet's volume, and $\mu_B$ = Bohr magneton. 

From the last three equations, we derive two coupled equations that describe the time-evolution 
of $\theta(t)$ and $\phi(t)$ (see the accompanying supplementary material \cite{supplement}).
After noise deflects the magnetization to a new orientation, we take that new orientation 
as the initial 
orientation [$\theta_{initial}, \phi_{initial}$] and solve the above coupled equations 
 numerically to determine the final state of the magnetization vector
[$\theta_{final}, \phi_{final}$]. The final state is reached when the orientation 
does not change further. 

In our simulations, the initial conditions span the space of
$\phi_{initial}$ from 0$^{\circ}$ to 360$^{\circ}$ (with four narrow excluded regions;
see later) and $\theta_{initial}$ from 85$^{\circ}$
 to 95$^{\circ}$. The angular resolution is 1$^{\circ}$. The results outside the interval 
$\left [ 85^{\circ} \leq \theta_{initial} \leq 95^{\circ} \right ]$
are not considered since they correspond to large out-of-plane excursions of the 
magnetization vector that can only occur when the noise amplitude is extremely high. 
Such large perturbations induce non-linear effects and will be discussed elsewhere.

The narrow regions excluded from the $\phi_{initial}$ space are the following. 
The states $\phi = 45^{\circ}, 135^{\circ}, 225^{\circ}$ and $315^{\circ}$ are the highest 
energy states when $\theta = 90^{\circ}$. If the magnetization is driven to any of these 
states by noise (unlikely since these are the highest energy states), 
it has equal probability of decaying to either one of its two neighboring minimum energy
(stable) states. Therefore, we exclude 
a region of 1$^{\circ}$ around these critical points.

In order to convey the results of magnetization dynamics obtained by solving the LLG equation, 
we use the following color scheme (see Fig. 1(b)): $blue$ corresponds to $\phi = 0^{\circ}$,
$red$ to $\phi = 90^{\circ}$,
$yellow$ to $\phi = 180^{\circ}$, and 
$green$ to $\phi = 270^{\circ}$.  
Fig. 1(c) illustrates the magnetization vector's final orientation 
for all allowed values of initial (perturbed) orientations $\theta_{initial}$ and 
$\phi_{initial}$. 
If $\theta_{initial}$ and $\phi_{initial}$ are in the blue region,
then the final state is always $\phi_{final} = 0^{\circ}$ and $\theta_{final}
 = 90^{\circ}$. Similarly, if $\theta_{initial}$ and $\phi_{initial}$ are in the red region,
then the final state is always $\phi_{final} = 90^{\circ}$ and $\theta_{final}
 = 90^{\circ}$, etc.
This plot shows that if $\theta_{initial} = 90^{\circ}$, then the final state for the 
range of initial states
$-44^{\circ} < \phi_{initial} < 44^{\circ}$ is $\phi_{final} = 0^{\circ}$
and $\theta_{final}
 = 90^{\circ}$.
Similarly, the final state for $\theta_{initial} = 90^{\circ}$ and the range of initial states
$46^{\circ} < \phi_{initial} < 134^{\circ}$ is $\phi_{final} = 90^{\circ}$ and 
$\theta_{final} = 90^{\circ}$, 
and so on. 
This means that if noise deflects the magnetization vector from a stable state while still 
staying in-plane, the vector always returns to the original state as long as the angular 
in-plane deflection is within $\pm 44^{\circ}$. 

However, when $\theta_{initial} = 89^{\circ}$, i.e. noise causes -1$^{\circ}$
out-of-plane excursion 
of the magnetization vector, the regions shift by -5$^{\circ}$. Now, if the magnetization's
initial state was in the interval [-$50^{\circ} \leq \phi_{initial} \leq 40^{\circ}$],
the the final state will be $\phi_{final} = 0^{\circ}$, and so on. This -5$^{\circ}$
shift happens because the out-of-plane excursion of the magnetization vector generates 
an additional precessional torque acting on the magnetization vector owing to the
coupled $\theta-\phi$ dynamics. Similarly, when $\theta_{initial} = 91^{\circ}$,
a shift of +5$^{\circ}$ takes place. As long as the out-of-plane excursion
$\Delta \theta_{initial} (= \theta_{initial} - 90^{\circ})$ is small, the shift is 
linearly proportional to $\Delta \theta_{initial}$. For $\Delta \theta_{initial} > 5^{\circ}$,
the shift increases non-linearly. Therefore, out-of-plane excursion is particularly harmful. 
As long as the magnetization vector is not deflected too far out of the magnet's plane,
and does not stray into the ambit of another stable state,
it always returns to the original state and recovers. 
This self-correcting behavior lends itself to image recovery. 

\begin{figure}

\includegraphics[width=2.5in]{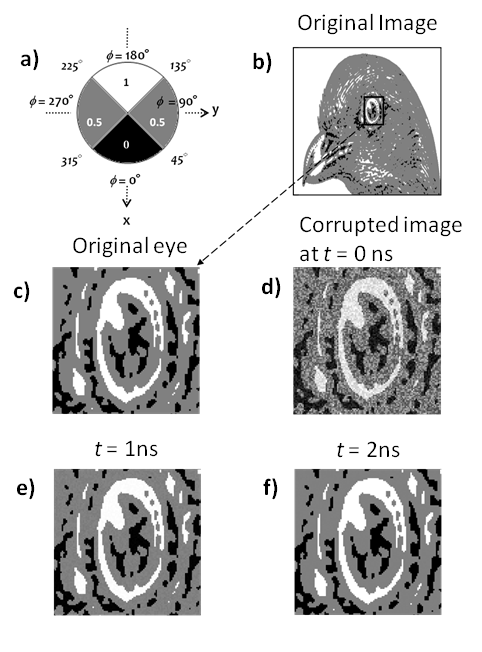}
\caption{(a) A pixel's shade is given a value of 0 (black), 0.5 (gray) or 1 (white). (b) Original 512$\times$512 
pixel image. Each pixel's 
shade is encoded into the magnetization state of the corresponding magnet. (c) The eye of the bird
 is blown up for clarity. (d) Noise corrupts the 
image at time $t$ = 0 by deflecting the magnetization vectors from their initial states. 
The out-of-plane excursion is restricted to 1$^{\circ}$. (e) The partially recovered image at $t$ = 1 ns. 
 (f) 
The fully recovered image at $t$ = 2 ns. 
 }
\end{figure}

Fig. 2 (a) shows the scheme for encoding pixel shades in a black/gray/white image. 
The magnetization's orientation encodes three different shades as follows: $\phi = 0^{\circ}$
 (black), $\phi = 90^{\circ}$ (gray), $\phi = 270^{\circ}$ (gray) and $\phi = 180^{\circ}$
  (white). The shades can also be assigned numerical values: black = 0, gray = 0.5 and 
  white = 1. 

A 512$\times$512 pixel image is encoded by the above scheme and is shown in Fig. 2 (b). 
Next, the numerical value of each pixel is changed randomly to simulate the effect of noise. 
We restrict the random out-of-plane deflection of
the magnetization vector $\Delta \theta$ to $\pm 1^{\circ}$, which then restricts the in-plane deflection 
$\Delta \phi$ to $\pm 40^{\circ}$
since that is the maximum in-plane deflection that can be corrected when $|\Delta \theta| \leq 1^{\circ}$.
The choice of $\pm 1^{\circ}$ out-of-plane deflection is dictated by the fact that this allows a 
reasonably large azimuthal deflection. This visually distorts the image by a large degree in Fig. 
2(d).
In accordance with this choice, a pixel with intensity value = 0 is randomly assigned a value between 0 and 
0.222. If the intensity value = 0.5, it is changed to something between 0.278 and 0.722. 
Similarly, for pixels with intensity value = 1, the value is changed to something between
 0.778 and 1. These distortions restrict the azimuthal deflections 
to $\pm 40^{\circ}$, i.e. $|\Delta \phi| \leq 40^{\circ}$.

Fig. 2 (c) shows the corrupted image at $t$ = 0 ns for a critical region (the bird's eye). By converting each pixel of the 
corrupted image to its equivalent $\phi$-value and then solving the coupled equations 
for $\theta-\phi$ dynamics \cite{supplement}, we can determine the final state (black, gray or white) of 
the pixel. Figs. 2 (d), (e) and (f) illustrate the image recovery process. At $t$ = 1 ns, the noise in the image has been greatly reduced, 
with most pixels settling back into their original states (Fig. 2 (e)). 
Steady state is achieved in 2 ns (Fig. 2 (f)) since the images at 2 ns and at 3 ns 
(the latter not shown here) are identical.
The final steady state image is identical 
with the original image pixel by pixel, showing 100\% recovery.

One could extend this scheme to pattern recognition as well. 
This would be achieved by integrating a magnetic tunnel junction (MTJ) vertically 
underneath each magnet. The magnetization orientations of the lower hard magnetic layers 
store the image to be compared against. The input image is written in the upper soft 
magnetic layers (Ni). If the input matches the stored image pixel by pixel, then the magnetizations 
of the soft and hard magnets will be all parallel. This will result in the maximum tunnel current 
flowing through each MTJ. By setting an appropriate current threshold (say X\% of 
the maximum), we can determine if the two images match with probability of X\% and thus 
recognize the input image. 

This work is supported by the US National Science Foundation under the Nanoelectronics 
Beyond the Year 2020  grant ECCS-1124714.


\end{document}